\documentclass[referee,useAMS,usenatbib,usegraphicx]{mn2e}
\usepackage{amssymb}
\usepackage{times}

\title[Gamma-ray Absorption in Cen~A]{Absorption of Nuclear Gamma-rays on the Starlight Radiation in FR~I Sources: the Case of Centaurus~A}

\author[Stawarz et al.]{\L . Stawarz$^{1, \, 2, \, 3,}$\thanks{E-mail:
Lukasz.Stawarz@mpi-hd.mpg.de}, F. Aharonian$^2$, S. Wagner$^1$, and M. Ostrowski$^3$\\
$^1$Landessternwarte Heidelberg, K\"onigstuhl, D-69117 Heidelberg, Germany\\
$^2$Max-Planck-Institut f\"ur Kernphysik, Saupfercheckweg 1, 69117 Heidelberg, Germany\\
$^3$Obserwatorium Astronomiczne, Uniwersytet Jagiello\'nski, ul. Orla 171, 30-244 Krak\'ow, Poland}

\begin{document}

\date{}

\pagerange{\pageref{firstpage}--\pageref{lastpage}} \pubyear{...}

\maketitle
\label{firstpage}

\begin{abstract}
Several BL Lac objects are confirmed sources of variable and strongly Doppler-boosted TeV emission produced in the nuclear portions of their relativistic jets. It is more than probable, that also many of the FR I radio galaxies, believed to be the parent population of BL Lacs, are TeV sources, for which Doppler-hidden nuclear $\gamma$-ray radiation may be only too weak to be directly observed. Here we show, however, that about one percent of the total time-averaged TeV radiation produced by the active nuclei of low-power FR~I radio sources is inevitably absorbed and re-processed by photon-photon annihilation on the starlight photon field, and the following emission of the created and quickly isotropized electron-positron pairs. In the case of the radio galaxy Centaurus~A, we found that the discussed mechanism can give a distinctive observable feature in the form of an isotropic $\gamma$-ray halo. It results from the electron-positron pairs injected to the interstellar medium of the inner parts of the elliptical host by the absorption process, and upscattering starlight radiation via the inverse-Compton process mainly to the GeV$-$TeV photon energy range. Such a galactic $\gamma$-ray halo is expected to possess a characteristic spectrum peaking at $\sim 0.1$ TeV photon energies, and the photon flux strong enough to be detected by modern Cherenkov Telescopes and, in the future, by GLAST. These findings should apply as well to the other nearby FR~I sources.
\end{abstract}


\section{Introduction}

Centaurus A is the most nearby active galaxy\footnote{Distance $d = 3.4$ Mpc; $1$ arcsec corresponds to $16$ pc.}, hosted by a powerful elliptical \citep[see a monograph by][]{isr98}. At low frequencies it reveals a giant ($8\deg \times 4\deg$ or $500$ kpc $\times$ $250$ kpc) and complex radio structure, with a total $5$ GHz energy flux of $3.4 \times 10^{-11}$ erg cm$^{-2}$ s$^{-1}$, containing a one-sided kpc-scale jet, few-kpc-long inner lobes, and extended outer lobes. The host galaxy is of $14' \times 18'$ optical size, has a blue luminosity of $7.5 \times 10^{10} \, L_{\odot}$, and contains a $\sim 10^8 \, M_{\odot}$ supermassive black hole in the center \citep{mar06}. The famous `dark lane' pronounced within the galactic body, being in fact an edge-on disk of rotating gas, dust and young stars, is most probably a remnant of the merger with a spiral galaxy, which happened some $10^8 - 10^9$ years ago. The kpc-scale jet and the active nucleus are confirmed sources of X-ray emission \citep[see most recent analysis by][and references therein]{har03,kat06}.

Cen~A source is classified as an FR~I radio galaxy, although some of its morphological characteristics (double-double structure, one-sided jet) are not typical for this class of AGNs. FR~I galaxies are believed to be a parent population of BL Lac objects \citep{urr95}, some of which are confirmed sources of very high energy (VHE) $\gamma$-ray emission. In fact, the nuclear part of the Cen~A jet was modeled in terms of a `misaligned' (and therefore Doppler-hidden) BL Lac, since the jet viewing angle in this source is expected to be much larger than the typical inclination of the blazar jet \citep[e.g.,][]{bai86,mor91,bot93,ste98,chi01}. As such, Cen~A was considered for some time as a potential source of TeV radiation, for which jet misalignment effects are compensated by the source's proximity. This expectation was strengthened by the positive detection of the considered object at MeV$-$GeV energy range by all the instruments on board CGRO during the period 1991-1995 \citep[and references therein]{ste98}. Nevertheless, the observations performed till now at the TeV energy range resulted in the upper limit for such an emission, only \citep{row99,aha04,aha05}.

In this paper we investigate the emission resulting from re-procession of the potential $\gamma$-ray flux of the active nucleus in the Cen~A radio galaxy via annihilation of the nuclear $\gamma$-rays on the starlight photon field due to the host galaxy, followed by the synchrotron and inverse-Compton cooling of the created electron-positron pairs. We show, that the analysis of this mechanism --- which also applies to other FR~I sources --- can give us interesting constraints on the unknown parameters of the Doppler-hidden and heavily obscured Cen~A nucleus, as well as of its host galaxy and outer parts of its radio outflow. We note that $\gamma$-ray opacity in blazar sources was considered before only in the context of nuclear target radiation fields, due to accretion discs, broad-line regions, or dusty tori \citep[e.g.,][]{sik94,der94,bla95,wag95,boe97,wan00}. In the case of the FR~I/BL Lac sources, however, such radiation fields are relatively weak, or even absent. Thus, the main source of the opacity for the TeV photons emitted from their active nuclei is the starlight radiation\footnote{We note, that ellipticals hosting radio sources may contain relatively large amounts of cold dust, prounounced at far infrared frequencies \citep{gol88,kna90}, for which the exact location is unknown. As mentioned before, the Cen~A system itself is bright in infrared frequencies due to the dust and young stars emission, restricted however to the `dust lane', i.e. relatively thin but extended disc-like feature (roughly perpendicular to the jet axis), being a remnant of the spiral which merged recently with the elliptical host. A role of this feature in absorbing and reprocessing the nuclear $\gamma$-rays is briefly discussed in Appendix A.}. This radiation can be modelled quite precisely, since all the low-power radio sources of the FR~I/BL Lac type are hosted by giant elliptical galaxies \citep[e.g.,][]{col95,urr00,hei04}, for which spectral and spatial distribution of the stellar emission is relatively well known \citep[see in this context discussion in][]{sta05,sta06a,sta06b}.

Below, in section 2, we present details regarding the calculation of the optical depth for photon-photon pair production by the Cen~A nuclear $\gamma$-ray emission on the starlight photon field of the elliptical galaxy, and the energetics of the created electron-positron pairs. In section 3, we discuss further evolution of such particles injected into the interstellar medium of the elliptical host, calculating in particular the resulting synchrotron and inverse-Compton fluxes. Section 4 contains a final discussion and conclusions.
\section{Calculations}

\subsection{Opacity}

In order to calculate the opacity for the $\gamma$-ray photon beam propagating through the galaxy, one has to specify the spectral and spatial distribution of the stellar photon field(s). Starlight surface brightness in elliptical galaxies, including those with active nuclei, are typicaly well fitted by an empirical Nuker law \citep{lau95}. In terms of monochromatic starlight intensity, this law can be expressed as
\begin{equation}
I(r) = I_{\rm b} \, 2^{(b - d)/a} \, \left({r \over r_{\rm b}}\right)^{-d} \, \left[1+\left({r \over r_{\rm b}}\right)^{a}\right]^{-(b-d)/a} \, ,
\end{equation}
\noindent
where $r$ is the distance from the galactic nucleus, $r_{\rm b}$ is the `break' radius, and $I_{\rm b} = I(r_{\rm b})$. Such profiles imply a power-law dependance $I(r) \propto r^{-d}$ for $r < r_{\rm b}$, and $I(r) \propto r^{-b}$ at larger distances. \citet{rui05} found that in the case of weak radio galaxies (selected from the B2 sample with the criterium of not showing significant dust emission), the host galaxies are characterized by the typical values of $a = 1.9$, $b = 1.6$, and $d = 0.02$. In a specific case of the Cen~A host galaxy, \citet{cap05} give $a = 1.68$, $b = 1.3$, $d = 0.1$, and a break radius $r_{\rm b} = 2.56\arcsec = 41$ pc. These values are considered hereafter, allowing to approximate the starlight emissivity, $j(r) \propto r^{-1} \, I(r)$, in the NGC~5128 (Cen~A host) galaxy, as
\begin{equation}
\epsilon j_{\epsilon}(\xi) = j_{\rm V} \, g(\epsilon) \, h(\xi) \, ,
\end{equation}
\noindent
where $\xi \equiv r / r_{\rm b}$. Here $\epsilon \equiv \varepsilon / m_{\rm e} c^2$ is the starlight photon energy $\varepsilon$ in $m_{\rm e}c^2$ units, $j_{\rm V}$ is the total $V$-band galactic emissivity, $g(\epsilon)$ describes the $V$-band normalized spectral distribution of the stellar photon field, and
\begin{equation}
h(\xi) = \left({r \over r_{\rm b}}\right)^{-1} \, {I(r) \over I_{\rm b}} = 1.64 \, \, \xi^{-1.1} \, \left(1+\xi^{1.68}\right)^{-0.7143}
\end{equation}
\noindent
is the radial depedance function. The considered galaxy is $14' \times 18'$ in optical size \citep{isr98}, and hence we take the average terminal galactic radius of $16'$, or $\xi_{\rm t} \equiv r_{\rm t} / r_{\rm b} = 375$, leading to the galactic volume (${\cal V}$) integral 
\begin{equation}
{\cal H} \equiv \int_{\cal V} h(\xi) \, d{\cal V} = 4 \pi \, r_{\rm b}^3 \int_0^{\xi_{\rm t}} \xi^2 \, h(\xi) \, d\xi = 1.83 \times 10^3 \, r_{\rm b}^3
\end{equation}
\noindent
(assuming spherical symmetry). For the spectral distribution of the NGC 5128 starlight, we assume the template spectrum of a powerful elliptical galaxy as provided by \citet{sil98}, restricted to the photon energy range between $\epsilon_{\rm min} = 10^{-7}$ and $\epsilon_{\rm max} = 10^{-5}$. This restriction implies that we consider only the direct stellar emission of the evolved red giants (constituting the main body of the elliptical host) and their winds, but not the far infrared emission resulting from the reprocession of the starlight photons by the cold galactic dust (see Appendix A). We normalize it to the $V$-band radiation, so that $g(\epsilon = h/m_{\rm e} c \lambda_{\rm V}) = 1$, where $\lambda_{\rm V} = 0.55$ $\mu$m. The resulting spectral function $g(\epsilon)$ is shown in Figure 1 (top-panel): values read directly from \citet{sil98} are represented by crosses, and the broken power-law approximation considered hereafter,
\begin{eqnarray}
g(\epsilon) = \left\{ \begin{array}{lll} 2.67 \times 10^{14} \, \epsilon^{2.44} & {\rm for} & 10^{-7.0} \leq \epsilon \leq 10^{-5.8} \\ 9.25 \times 10^{-4} \, \epsilon^{-0.57} & {\rm for} & 10^{-5.8} < \epsilon < 10^{-5.3} \\ 3.54 \times 10^{-31} \, \epsilon^{-5.74} & {\rm for} & 10^{-5.3} \leq \epsilon \leq 10^{-5.0} \end{array} \right. ,
\end{eqnarray}
\noindent
is shown as a solid line.

\begin{figure}
\includegraphics[scale=1.50]{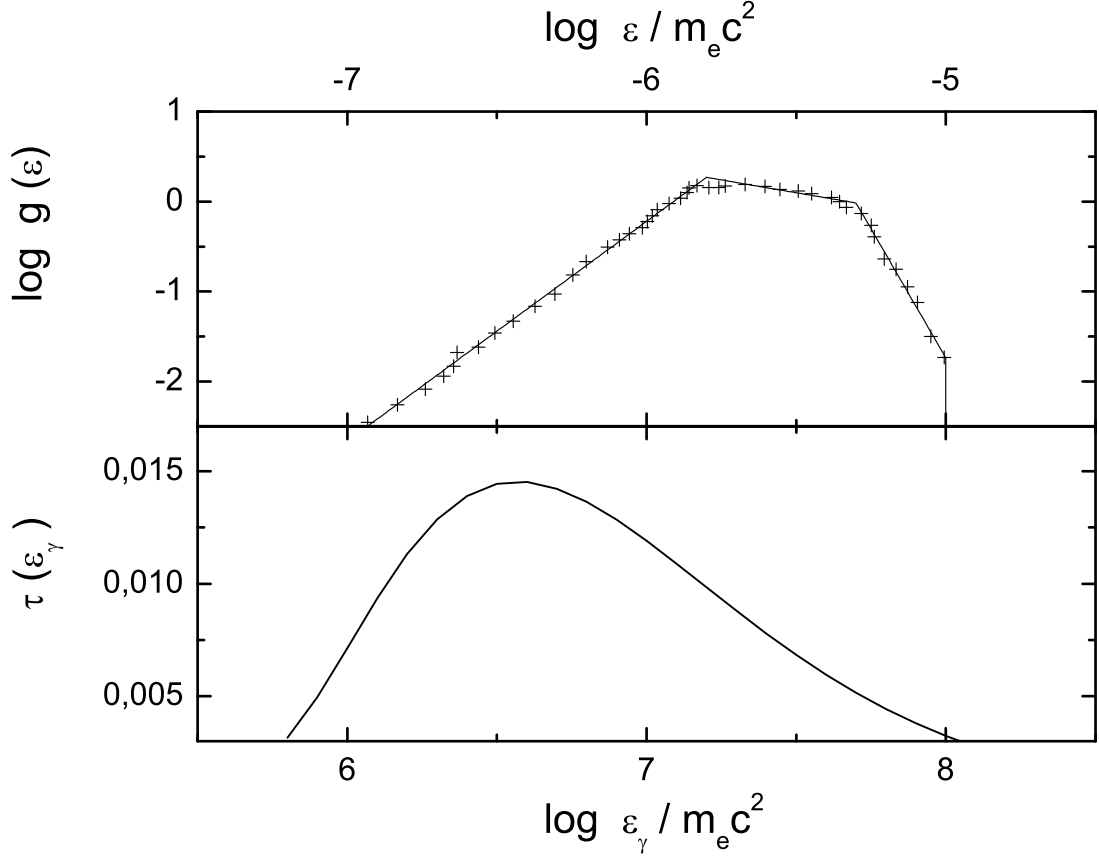}
\caption{{\it Top:} Spectral energy distribution of the starligh emission of a template giant elliptical; values read directly from \citet{sil98} are represented by crosses, and the broken power-law approximation introduced in this paper is shown as a solid line. {\it Bottom:} Optical depth for annihilation of the nuclear $\gamma$-ray emission on the starlight photon field.}
\end{figure}

The apparent $V$-band magnitude of NGC 5128 is $m_{\rm V} = 6.98$ \citep{isr98}. This gives the monochromatic $V$-band galactic luminosity
\begin{equation}
\log \left({L_{\rm V} \over {\rm erg/s}}\right) = 50.078 + 2 \, \log \left({d_{\rm L} \over {\rm Mpc}}\right) -0.4 \, (m_{\rm V} - A_{\rm V}) - c_{\rm V} = 43.82 \quad ,
\end{equation}
\noindent  
where the distance of Cen~A is $d_{\rm L} = 3.4$ Mpc, the extinction is $A_{\rm V}=0.381$, and $c_{\rm V} = 4.68$. Since, by the definition,
\begin{equation}
L_{\rm V} = 4 \pi \, \int_{\cal V} \left[\epsilon j_{\epsilon}(\xi)\right]_{\rm V} \, d{\cal V} = 4 \pi \, j_{\rm V} \, {\cal H} \quad ,
\end{equation}
\noindent  
one obtains $j_{\rm V} = 1.42 \times 10^{-21}$ erg cm$^{-3}$ ster$^{-1}$ s$^{-1}$. Note, that by means of standard radiative transfer formulae, the spectral intensity within a solid angle $\Omega$, which can be obtained by integrating spectral emissivity along a ray as $I_{\nu} (\Omega) = \int \, j_{\nu}(\xi) \, dl$, is related to the differential photon number density, $n_{\epsilon}(\xi, \Omega)$, through the expression $I_{\nu}(\Omega) = c h \, \epsilon \, n_{\epsilon}(\xi, \Omega)$. This corresponds in fact to the neglegible absorption of the starlight along the light path, the approximation which may be considered as being in conflict with the presence of the dust lane in Cyg~A host. It is however good enough for the purpose of the presented evaluation, since integration over the whole extended elliptical volume reduces the starlight absorption effects. As a result,
\begin{equation}
n_{\epsilon}(\xi, \Omega) = {\epsilon^{-2} \over m_{\rm e} c^3} \, \int \, \left[\epsilon j_{\epsilon}(\xi)\right] \, dl \quad .
\end{equation}
\noindent

The optical depth for photon-photon annihilation, computed for the case of a monodirectional beam of $\gamma$-ray photons with a dimensionless energy $\epsilon_{\gamma}$, propagating through the stellar photon field of the host galaxy from the active center up to the terminal distance $r_{\rm t}$, is
\begin{equation}
\tau(\epsilon_{\gamma}) = \int_0^{r_{\rm t}} dr \, \int \, dn \, (1-\varpi) \, \sigma_{\gamma \gamma} \quad ,
\end{equation}
\noindent
where $\varpi$ is the cos function of the angle between the $\gamma$-ray photon and the incident starlight photon, $dn = n_{\epsilon}(\xi, \Omega) \, d\epsilon \, d\Omega$ is the differential starlight photon number density, and
\begin{equation}
\sigma_{\gamma \gamma}(\epsilon_{\gamma}, \epsilon, \varpi) = {3 \sigma_{\rm T} \over 16} \, (1 - \beta^2) \, \left[ (3 - \beta^4) \, \ln \left( {1 + \beta \over 1 - \beta} \right) - 2 \beta \, (2 - \beta^2) \right]
\end{equation}
\noindent
is the photon-photon annihilation cross section \citep{gou67}, where
\begin{equation}
\beta \equiv \left( 1 - {2 \over \epsilon_{\gamma} \epsilon \, ( 1- \varpi)} \right)^{1/2} \quad ,
\end{equation}
\noindent
is the velocity of the created electron/positron (computed via the conservation of four-momentum) in the appropriate center-of-momentum frame. By choosing $d\Omega = d\phi \, d\varpi$, the differential starlight photon number density (see equation 8) reads as
\begin{equation}
n_{\epsilon}(\xi, \Omega) = {\epsilon^{-2} \, r_{\rm b} \over m_{\rm e} c^3} \, \int_0^{\eta_{\rm max}} \, \left[\epsilon j_{\epsilon}(\zeta)\right] \, d\eta \quad ,
\end{equation}
\noindent
where $\eta \equiv l / r_{\rm b}$, $\zeta = \sqrt{\xi^2 + \eta^2 - 2 \xi \eta \varpi}$, and the integration upper limit is $\eta_{\rm max} = \xi \varpi + \sqrt{\xi^2 \varpi^2 - \xi^2 + \xi_{\rm t}^2}$. This gives finaly
\begin{equation}
\tau(\epsilon_{\gamma}) = {2 \pi \, j_{\rm V} \, r_{\rm b}^2 \over m_{\rm e} \, c^3} \, \int_{-1}^{+1} d\varpi \, (1 - \varpi) \, \int_{\max(\epsilon_{\rm min}, \epsilon_{\rm thre})}^{\epsilon_{\rm max}} d\epsilon \, \sigma_{\gamma \gamma}(\epsilon_{\gamma}, \epsilon, \varpi) \, { g(\epsilon) \over \epsilon^2 } \, \int_0^{\xi_{\rm t}} d \xi \int_0^{\eta_{\rm max}} d\eta \, h(\zeta) \, ,
\end{equation}
\noindent
where the threshold energy is $\epsilon_{\rm thre} = 2 / \epsilon_{\gamma} \, (1 - \varpi)$. This optical depth as a function of the $\gamma$-ray photon energy is shown in Figure 1 (bottom-panel). As can be seen, it possesses a broad maximum at $\varepsilon_{\gamma} \approx 1-10$ TeV, with the maximum value $\tau(2 \, {\rm TeV}) \approx 0.0145$.

Let us fix the $\gamma$-ray photon energy at $\epsilon_{\gamma} = 10^{6.6}$, and compute again the appropriate optical depth for photon-photon annihilation as a function of the upper bound in integration over the distance from the core, $\tau = \tau(\xi)$, replacing simply constant $\xi_{\rm t}$ in equation 13 with variable $\xi$. Such a profile is shown in Figure 2 (solid line): as can be seen, only about $40\%$ ($\tau \approx 0.006$) of the annihilated radiation is in fact absorbed within the innermost galactic volume of radius $r_{\rm b}$, and the remaining $60\%$ is absorbed at distances $r_{\rm b} < r < 100 \, r_{\rm b}$. The obtained profile for $\tau$ can be compared with the spatial distribution of the galactic radiation. Noting the definition for the photon field spectral energy density,
\begin{equation}
U_{\nu} = {2 \pi \over c} \, \int_{-1}^{+1} I_{\nu}(\Omega) \, d\varpi \quad ,
\end{equation}
\noindent
one may find the appropriate bolometric energy density profile for the starlight emission as
\begin{equation}
U_{\rm rad}(\xi) = f_{\rm bol} \, \left[\nu U_{\nu}\right]_{\rm V} = f_{\rm bol} \, {2 \pi \, r_{\rm b} \, j_{\rm V} \over c} \, \int_{-1}^{+1} d\varpi \, \int_0^{\eta_{\rm max}} d \eta \, \, h(\zeta) \quad ,
\end{equation}
\noindent
where $f_{\rm bol} = 2.5$ is the $V$-band bolometric correction for the adopted here stellar spectrum (equation 5). This profile is shown on Figure 2 as a dotted line.

\begin{figure}
\includegraphics[scale=1.50]{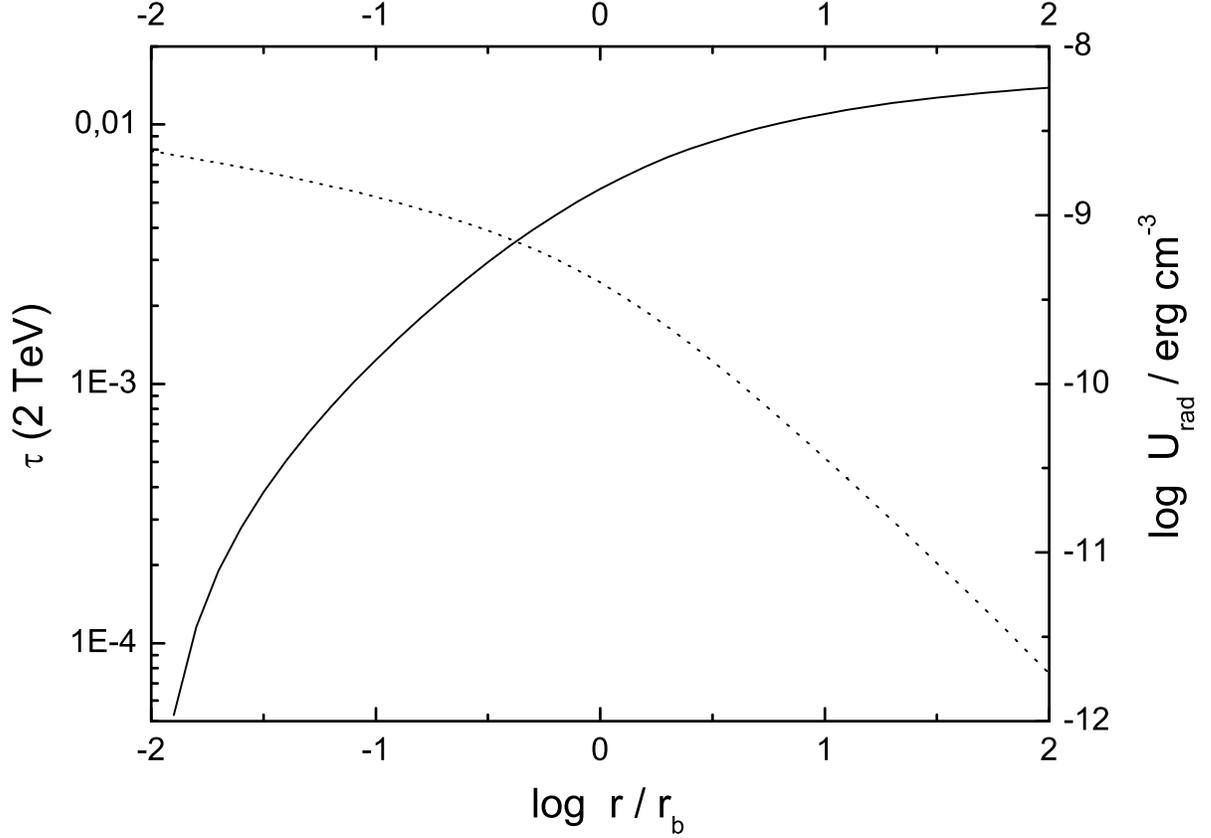}
\caption{{\it Left:} Optical depth for $2$ TeV photons as a function of the distance from the active nucleus (solid line). {\it Right:} Profile of the energy denisty of the starlight photon field (dotted line).}
\end{figure}

The spectral and spatial shapes of the optical depth for the photon-photon annihilation discussed above for the Cen~A radio galaxy is universal, and should hold also for the other low-power radio sources hosted by elliptical galaxies. That is caused by the appropriate scaling $\tau \propto j_{\rm V} \, r_{\rm b}^2$ (equation 13), $j_{\rm V} \propto L_{\rm V} / {\cal H}$ (equation 7), and ${\cal H} = \kappa \, r_{\rm b}^3$ (equation 4), where $\kappa = \kappa(a, b, d)$ is a numerical factor depending on the parameters of the Nuker profile (for example, $\kappa = 1.83 \times 10^3$ in the case of NGC~5128, as given in the equation 3). These relations give $\tau \propto L_{\rm V} / \kappa \, r_{b}$. Since the parameters of the NGC~5128 Nuker profile \citep[$a = 1.68$, $b = 1.3$, $d = 0.1$][]{cap05} are similar to the average hosts of B2 radio sources \citep[$a = 1.9$, $b = 1.6$, $d = 0.02$][]{rui05}, since the spectral energy distribution of the starlight emission considered here is based on a universal spectrum of the elliptical galaxy \citep{sil98}, and since the break radius scales linearily with the $V$-band galactic luminosity,
\begin{equation}
\left({r_{\rm b} \over {\rm kpc}}\right) \approx \left({L_{\rm V} \over 10^{45} \, {\rm erg/s}}\right) \, , 
\end{equation}
\noindent
\citep{rui05}, the optical depth for the $\gamma$-ray photons shown on Figures 1 and 2 for the case of Cen~A radio galaxy indeed should apply also to the other analogous sources. We note for example, that with the value of $L_{\rm V}$ found previously for the NGC~5128 galaxy, the simple scaling introduced in equation 16 implies the break radius $\approx 66$ pc, which is very close to the observed value of $r_{\rm b} \approx 41$ pc \citep{cap05}. On the other hand, the starlight energy density profile $U_{\rm rad} \propto j_{\rm V} \, r_{\rm b} \propto r_{\rm b}^{-1}$ is supposed to change for different sizes (and hence luminosities) of the elliptical host when compared with the particular profile shown on Figure 2 for the case of NGC~5128 parameters. 

\subsection{Energetics}

The obvious source of $\gamma$-ray photons in the Cen~A system is the active nucleus (containing a relativistic jet), which is most probably responsible for production of the flux detected by OSSE ($0.05-4$ MeV), COMPTEL ($0.75-30$ MeV) and EGRET ($0.1-1.0$ GeV) instruments on board CGRO in the period 1991 -- 1995, with the total observed $50$ keV -- $1$ GeV luminosity $\sim 3 \times 10^{42}$ erg s$^{-1}$ \citep[and references therein]{ste98}. This emission peaks at $\approx 0.1$ MeV, with the maximum energy flux about $\sim 10^{-9}$ erg cm$^{-2}$ s$^{-1}$. At a very high $\gamma$-ray energy range, a 3$\sigma$ detection of Cen~A with the photon flux $F(>0.3 \, {\rm TeV}) = 4.4 \times 10^{-11}$ ph cm$^{-2}$ s$^{-1}$ was reported in the seventies \citep{gri75}. Subsequent CANGAROO observations resulted in the upper limit $F(>1.5 \, {\rm TeV}) < 5.45 \times 10^{-12}$ ph cm$^{-2}$ s$^{-1}$ for the point source centered at the Cen~A nucleus, and $F(>1.5 \, {\rm TeV}) < 1.28 \times 10^{-11}$ ph cm$^{-2}$ s$^{-1}$ for the extended region with radius $14'$ \citep{row99}. The most recent HESS observations gave $F(>0.19 \, {\rm TeV}) < 5.68 \times 10^{-12}$ ph cm$^{-2}$ s$^{-1}$ for the point-like active nucleus \citep{aha05}.

Here we are interested in the total, time-averaged and angle-averaged (i.e., `calorimetric') flux of TeV photons produced by the active nucleus and `injected' into the host galaxy as well as into the large-scale radio structure. With the most recent HESS results, we can only put some upper limits for it. Namely, assuming a standard apectral index $\alpha_{\gamma} = 1$ (equivalent to the photon index $\Gamma_{\gamma} = \alpha_{\gamma} + 1 = 2$) at the considered photon energy range, the relation between the integrated photon flux and the monochromatic flux energy density at any photon energy $\varepsilon \geq \varepsilon_0$ is simply $\left[\varepsilon S_{\varepsilon}\right] = \left[\varepsilon_0 \, F(>\varepsilon_0)\right]$. This flux is related to the emitting fluid (jet) intrinsic monochromatic power (assumed to be isotropic in the jet comoving frame) radiated in a given direction, $\partial L' / \partial \Omega' = L' / 4 \pi$, by the expression 
\begin{equation}
\left[\nu S_{\nu}\right] = {1 \over d_{\rm L}^2} \, {\delta_{\rm nuc}^3 \over \Gamma_{\rm nuc}} \, {\partial L' \over \partial \Omega'} = {1 \over d_{\rm L}^2} \, {\delta_{\rm nuc}^3 \over \Gamma_{\rm nuc}} \, {L' \over 4 \pi} \quad ,
\end{equation}
\noindent
where $\Gamma_{\rm nuc}$ and $\delta_{\rm nuc} = \Gamma_{\rm nuc}^{-1} \, \left(1 - \sqrt{1 - \Gamma_{\rm nuc}^{-2}} \cos \theta\right)^{-1}$ are, respectively, Lorentz and Doppler factors of the nuclear portion of the jet, and $\theta$ is the jet viewing angle \citep[e.g.,][]{sik97,sta03}. On the other hand, the total power radiated into the ambient medium being of interest here, is
\begin{equation}
L_{\rm inj} = \oint {\delta_{\rm nuc}^3 \over \Gamma_{\rm nuc}} \, {\partial L' \over \partial \Omega'} \, d\Omega = {1 \over 2} \, L' \, \Gamma_{\rm nuc}^{-1} \, \int_0^{\pi} \, \delta_{\rm nuc}^{-3} \, \sin \theta \, d\theta = L' \quad ,
\end{equation}
\noindent
and hence
\begin{equation}
L_{\rm inj} = 4 \pi \, d_{\rm L}^2 \, \Gamma_{\rm nuc} \delta_{\rm nuc}^{-3} \, \left[\varepsilon_0 \, F(> \varepsilon_0)\right] \quad . 
\end{equation}
\noindent
With the HESS photon flux $F(> 0.19 \, {\rm TeV}) < 5.68 \times 10^{-12}$ ph cm$^{-2}$ s$^{-1}$ (corresponding to the few-arcmin-integration area centered on the Cen~A nucleus) one obtains the upper limit for the total injected monochromatic power $L_{\rm inj} < 2.4 \times 10^{39} \, \Gamma_{\rm nuc} \delta_{\rm nuc}^{-3}$ erg s$^{-1}$. With the prefered values $\theta \sim 50\deg-80\deg$ inferred from the VLBI radio observations \citep{jon96,tin98}, and $\Gamma_{\rm nuc} \sim 10$ widely considered as a typical value for the bulk Lorentz factor of sub-parsec scale AGN jets, this reads as $L_{\rm inj} < 10^{42} - 10^{43}$ erg s$^{-1}$. Below we take conservatively $L_{\rm inj} = 10^{42}$ erg s$^{-1}$ as an upper limit, noting that such a luminosity would correspond to less than $10 \%$ of the minimum total kinetic power of the Cen~A jet, estimated to be $L_{\rm j} \gtrsim 10^{43}$ erg s$^{-1}$ from the radio/X-ray lobes' energetic \citep[see][]{cla92,kra03}. As shown in the previous section, about $1\%$ of this power is absorbed on the starlight photon field within the central $100 \, r_{\rm b} \sim 4$ kpc region of the host galaxy, and is thus converted to an electron-positron population, mainly in the $0.1-1$ TeV energy range.

The nuclear $\gamma$-ray emission postulated here is expected to be Doppler-boosted within the narrow cone characterized by the opening angle $\Gamma_{\rm nuc}^{-1} \lesssim 6\deg$. Therefore, it is strongly Doppler hidden when viewed from $\theta \geq 50\deg$ (see equation 17). If the observer was located within the beaming cone of this emission however, he would detect a flux corresponding to the isotropic luminosity $L(0) = (\delta_{\rm nuc, \, \theta=0}^3 / \Gamma_{\rm nuc}) \, L' \approx \Gamma_{\rm nuc}^2 \, L' < 10^{44}$ erg s$^{-1}$. Such values are consistent with luminosities observed from the TeV-detected BL Lac objects \citep[see, e.g.,][]{katar06}, which are believed to be beamed analogues of the low-power FR I radio galaxies like Cen~A \citep{urr95}.

\begin{figure}
\includegraphics[scale=1.50]{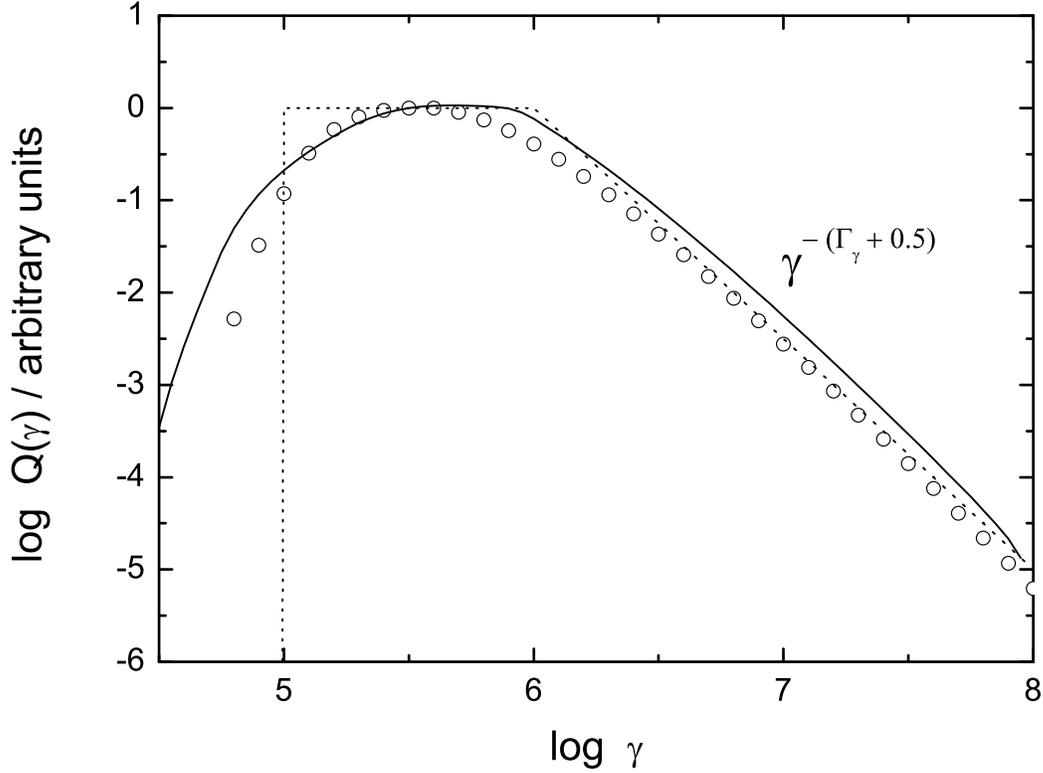}
\caption{Different approximations for the pair injection energy spectrum (corresponding to the photon flux of the primary $\gamma$-rays $\propto \epsilon_{\gamma}^{-2}$) in normalized units: (i) $Q(\gamma) \propto \gamma^{-2} \, \tau( 2 \gamma)$ (open circles), (ii) $Q(\gamma)$ calculated using the pair production rate expression introduced by \citet[solid line]{aha83}, and (iii) broken power-law approximation introduced in this paper (dotted line).}
\end{figure}

In order to find the energy spectrum of the created electrons and positrons, we note that the pair production rate for small values of the optical depth can be estimated as
\begin{equation}
Q(\gamma, r) \propto \left. \left\{n_{\gamma}(\epsilon_{\gamma}) \times {d \tau (\epsilon_{\gamma}) \over dr}\right\}\right|_{\epsilon_{\gamma} = 2 \, \gamma}
\end{equation}
\noindent
\citep[e.g.,][]{cop90,bot97}, where $\gamma$ is the Lorentz factor of the created particles, and $n_{\gamma}(\epsilon_{\gamma})$ is the photon spectrum of the primary $\gamma$-ray photons. For the latter we assume power-law form $n_{\gamma}(\epsilon_{\gamma}) \propto \epsilon_{\gamma}^{-\Gamma_{\gamma}}$, where $\Gamma_{\gamma} = \alpha_{\gamma} + 1$ is the photon index. When averaged over the galactic radius, the pair production rate therefore scales as
\begin{equation}
Q(\gamma) \propto \gamma^{-\Gamma_{\gamma}} \, \tau\left(2 \, \gamma\right) \quad .
\end{equation}
\noindent
This function is shown in Figure 3 in normalized units (open circles) for $\Gamma_{\gamma} = 2$, taken in this paper as a typical photon index of blazars' TeV emission. As illustrated, $Q(\gamma)$ is peaked at $\gamma \sim 10^{5.5}$, and decreases as $\propto \gamma^{- (\Gamma_{\gamma}+0.5)}$ for $\gamma > 10^6$. Such behavior is in fact expected, since for $\epsilon_{\gamma} > 10^{6.6}$ the optical depth evaluated previously is roughly $\tau(\epsilon_{\gamma}) \propto \epsilon_{\gamma}^{-0.5}$. Figure 3 shows also the pair injection energy distribution calculated using the expression introduced by \citet[solid line]{aha83}, integrated over the spatial-averaged spectrum of the starlight photons (equation 12; see also equation 34 below) and the assumed photon spectrum of primary $\gamma$-rays, $\propto \epsilon_{\gamma}^{-2}$. We note that the approximation of \citet{aha83} for the energy distribution of the secondary electrons provides accuracy better than $few \%$ \citep[see detailed calculations in][]{bot97}. Small differences between the two presented estimates come from the fact that the function given by \citet{aha83} corresponds to the isotropic distribution of the soft photons, while the optical depth calculated by us (i.e. equations 20-21) includes the anisotropy of the starlight radiation field. Finally, Figure 3 shows also the simplest broken power-law approximation for the injection pair spectrum considered hereafter for the purpose of the following calculations: $Q(\gamma) \propto const$ for $\gamma_0 \equiv 10^5 \leq \gamma \leq \gamma_{\rm br} \equiv 10^6$, and $Q(\gamma)\propto \gamma^{-2.5}$ for $\gamma > \gamma_{\rm br}$. We normalize it to the monochromatic $1$ TeV power $L_{\rm inj}$ specified above, to obtain
\begin{equation}
Q(\gamma) = {\tau \, L_{\rm inj} \, {\cal{I}}(\gamma) \over \gamma_{\rm br}^2 \, m_{\rm e} c^2 \, \cal{V}_{\rm gal}} \quad {\rm where} \quad {\cal{I}}(\gamma) = \left\{ \begin{array}{lll} 1 & {\rm for} & \gamma_0 \leq \gamma \leq \gamma_{\rm br} \\ (\gamma_{\rm br} / \gamma)^{2.5} & {\rm for} & \gamma > \gamma_{\rm br} \end{array} \right.
\end{equation}
\noindent
and zero otherwise. Here $\tau \equiv \tau(\epsilon_{\gamma} = 1 \, {\rm TeV}) \approx 0.01$, and $\cal{V}_{\rm gal}$ is the galactic volume to which the pair injection is taking place.

\section{Implications}

The electrons and positrons produced by annihilation of the VHE $\gamma$-ray nuclear photons on starlight radiation are expected to have an initial spectrum peaked at $0.1-1$ TeV energies. After being injected into the galactic medium, they are quickly isotropized by the ambient magnetic field, and radiate via the synchrotron and inverse-Compton processes. The evolution of these electrons and details of their emission spectra depend therefore on the properties of the interstellar medium of the elliptical host.\footnote{We also note, that since the nuclear $\gamma$-ray emission is expected to be beamed within the narrow cone with an opening angle of less then $10\deg$, as discussed in the previous section, some part of it illuminates also the large-scale radio outflow, i.e. the kpc-scale jet (extending in the case of Cen~A from $\sim 0.2$ kpc up to $\sim 4$ kpc from the center), and the `inner' lobe \citep[$\sim 4-5$ kpc; see][]{bur83,cla92}. The evolution of the electron-positron pairs injected via the photon-photon annihilation into these regions can differ substantially from the evolution of the particles injected into the body of the elliptical host as considered below. This problem will be investigated in a subsequent paper.}

\subsection{Elliptical host}

\citet{mos96} argued, that elliptical galaxies have no ordered large-scale magnetic field, but only an unresolved random component. They further suggested, that the latter is due to a `fluctuation dynamo' driven by the turbulent motions of the interstellar matter caused by type I supernovae and stellar motions. The latter ones are expected to be characterized by the Kolmogorov-like energy spectrum and an injection scale $\sim 3$ pc, while the former ones by a steeper (shock-like) spectrum and larger injection scale, $\sim 300$ pc. The resulting random magnetic field is expected to be characterized by the average galactic intensity $\sim 3$ $\mu$G (reaching $\sim 10$ $\mu$G in the central parts) and the correlation scale $\sim 100$ pc. \citet{mos96} demonstrated that their model is consistent with all the observational constraints (in particular with the depolarization studies). In the case of Centaurus A, and in general all the ellipticals hosting radio-loud AGNs, a galactic magnetic field can be even higher than this, especially close to the galactic center, possessing even some regular component due to polution of the interstellar medium by the magnetized plasma transported from the active core in the form of the jets. However, for the purpose of order-of-magnitude evaluations, below we take the characteristic values for the NGC~5128 elliptical host's magnetic field $B_{\rm gal} \approx 3-10$ $\mu$G, assuming that it consists solely of the (Alfv\'{e}nic) turbulent component with the maximum wavelength $\lambda_{\rm max} \sim 100$ pc and Kolmogorov energy spectrum $W(k) \propto k^{-q}$, where $q = 5/3$.

\begin{figure}
\includegraphics[scale=1.50]{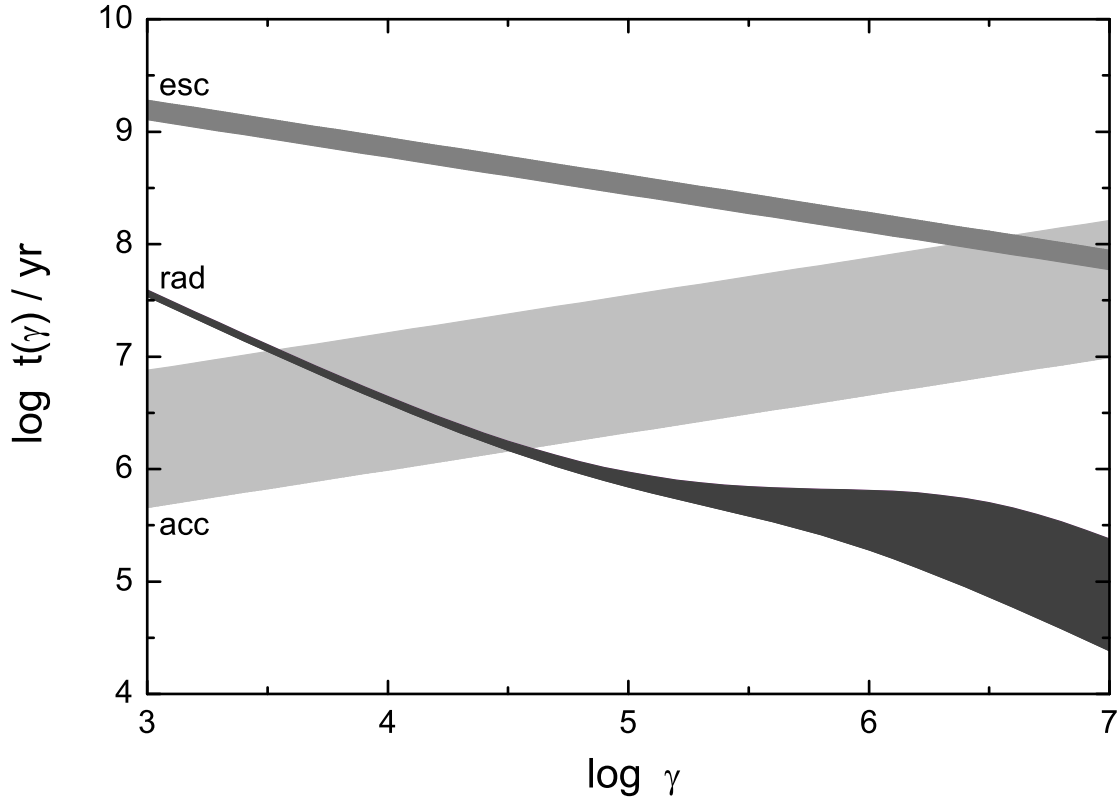}
\caption{Time scales for the radiative cooling (dark gray), turbulent re-acceleration (light gray) and diffusive escape (gray) of the electrons injected into the interstellar medium of the elliptical host, for the range of the galactic magnetic field $3-10$ $\mu$G. The values corresponding to $B_{\rm gal} = 10$ $\mu$G correspond to the lower bounds for the radiative losses and the acceleration time scales, and to the upper bound for the escape time scales, respectively.}
\end{figure}

With the interstellar medium parameters as discussed above, the mean free path of the created electron-positron pairs for resonant interactions with the turbulent Alfv\`{e}n modes, is \citep{sch89}
\begin{equation}
\lambda_{\rm e} \approx r_{\rm g} \, \left({\lambda_{\rm max} \over r_{\rm g}}\right)^{q-1} = r_{\rm g}^{1/3} \, \lambda_{\rm max}^{2/3} \sim 0.82 \times \gamma_6^{1/3} \, B_{-5}^{-1/3} \, {\rm pc} \quad ,
\end{equation}
where $\gamma_6 \equiv \gamma / 10^6$, $r_{\rm g} \equiv \gamma \, m_{\rm e} c^2 / e B_{\rm gal} \sim 5.5 \times 10^{-5} \ \gamma_6 \, B_{-5}^{-1}$ pc is the electrons' gyroradius, and $B_{-5} \equiv B_{\rm gal} / 10$ $\mu$G \citep[as a general reference for the particle-turbulent wave interactions see][]{sch02}. Such interactions lead to quick isotropisation of the injected electrons. In particular, the appropriate isotropisation time scale, $t_{\rm iso} \sim 3 \, \lambda_{\rm e} / c \sim 8 \, \gamma_6^{1/3} \, B_{-5}^{-1/3}$ yrs, is a few orders of magnitude shorter than the radiative cooling time scale (see below). The diffusive escape time scale from the central parts of the elliptical host ($R \sim 100 \, r_{\rm b} \sim 4$ kpc) is $t_{\rm esc} \sim 3 \, R^2 / \lambda_{\rm e} \, c \sim 2 \times 10^8 \, \gamma_6^{-1/3} \, B_{-5}^{1/3}$ yrs. We note, that the re-acceleration of the radiating particles by resonant scattering on the Alfv\`{e}n modes is not expected to be efficient enough to keep the electrons around $\gamma \sim 10^6$. In particular, the time scale for this process, $t_{\rm acc} \sim \beta_{\rm A}^{-2} \, t_{\rm iso} \sim 4.5 \times 10^6 \, \gamma_6^{1/3} \, B_{-5}^{-7/3}$ yrs, is much longer than the radiative cooling time scale. In the above, $v_{\rm A} \equiv \beta_{\rm A} c \approx B_{\rm gal} \, (4 \pi \, m_{\rm p} \, n_{\rm gas})^{-1/2} \sim 10^{-3} \, B_{-5} \, c$ is the Alfv\`{e}n velocity expected for the average number density of the cold gas within central ($< 4$ kpc) parts of the NGC~5128 galaxy $n_{\rm gas} \sim 3 \times 10^{-3}$ cm$^{-3}$. In this context we note that the most recent analysis presented by \citet{kra03} indicates that the hot gaseous component of the galaxy discussed here is characterized by a tempreature $kT \sim 0.3$ keV and a central number density $\sim 4 \times 10^{-2}$ cm$^{-3}$, which is roughly constant within the radius $r \sim 0.5$ kpc and decreases further away as $\propto r^{-1.2}$. Such a behavior is consistent with general properties of the elliptical galaxies \citep{mat03}. Thus, one can conclude that the TeV energy electrons injected into the interstellar medium (via annihilation of the $\gamma$-ray emission of the active center) are effectively confined to the elliptical body and quickly isotropized by the galactic magnetic field, and therefore radiate all their energy there by inverse-Compton upscattering of the starlight photons and the synchrotron process (see below), before being re-accelerated by the turbulent processes. As a result, one should expect formation of a small-scale (galactic) version of the `isotropic pair halos' discussed by \citet{aha94}, restricted however to the first generation of the secondary photons.

Below we investigate in more detail the emission spectrum of electron-positron pairs created within the centaral $100 \, r_{\rm b}$ parts of the elliptical host's interstellar medium due to annihilation of the nuclear VHE $\gamma$-rays on the starlight photon field. First, we introduce the \emph{averaged} energy density of the starlight photon field, assumed in this section to be isotropic within $100 \, r_{\rm b}$,
\begin{equation}
\epsilon U_{\epsilon} \approx \langle\left[\nu U_{\nu}\right]_{\rm V}\rangle_{\xi} \times g(\epsilon)
\end{equation}
\noindent
(see equations 5 and 15), where 
\begin{equation}
\langle\left[\nu U_{\nu}\right]_{\rm V}\rangle_{\xi} \equiv {\int_0^{\xi_{\rm cr}} d\xi \, \left[\nu U_{\nu}\right]_{\rm V} \over \int_0^{\xi_{\rm cr}} d\xi} = {2 \pi \, r_{\rm b} \, j_{\rm V} \over c \, \xi_{\rm cr}} \, \int_0^{\xi_{\rm cr}} \, d\xi \, \int_{-1}^{+1} d\varpi \, \int_0^{\eta_{\rm max}} d \eta \, \, h(\zeta) \quad ,
\end{equation}
\noindent
and $r_{\rm cr} \equiv \xi_{\rm cr} \, r_{\rm b}$ is some particular critical radius over which the averaging is performed. The total averaged starlight energy density is
\begin{equation}
\langle U_{\rm rad} \rangle = \int_{\epsilon_{\rm min}}^{\epsilon_{\rm max}} \, U_{\epsilon} \, d\epsilon = f_{\rm bol} \, \langle\left[\nu U_{\nu}\right]_{\rm V}\rangle_{\xi} \quad .
\end{equation}
\noindent
With $\xi_{\rm cr} = 100$ and other parameters as given above, one obtains $\langle\left[\nu U_{\nu}\right]_{\rm V}\rangle_{\xi} \approx 10^{-11}$ erg cm$^{-3}$, and $\langle U_{\rm rad} \rangle \approx 2.5 \times 10^{-11}$ erg cm$^{-3}$, which is still much higher than the energy density of the galactic magnetic field, $U_{\rm B} = B_{\rm gal}^2 / 8 \pi \approx 4 \times 10^{-12} \, B_{-5}^2$ erg cm$^{-3}$. The time-scale for the radiative losses of the electrons with Lorentz factor $\gamma$, including synchrotron and inverse-Compton losses (in both Thomson and Klein-Nishina regimes) can be then estimated as
\begin{equation}
t_{\rm rad}(\gamma) = {3 \, m_{\rm e} c \over 4 \, \sigma_{\rm T} \, U_{\rm B} \, \gamma \, (1 + q \, F_{\rm KN})} \quad ,
\end{equation}
\noindent
where $q \equiv \langle U_{\rm rad} \rangle / U_{\rm B} \approx 7 \, B_{-5}^{-2}$, and 
\begin{equation}
F_{\rm KN} = {1 \over \langle U_{\rm rad} \rangle} \, \int {U_{\epsilon} \over \left(1+ 4 \, \gamma \, \epsilon\right)^{1.5} } \, d\epsilon
\end{equation}
\noindent
\citep{mod05}. The simplified form of the function $F_{\rm KN}$ introduced above is in fact a very accurate approximation for the values $4 \, \gamma \, \epsilon < 4 \, \gamma_{\rm inj} \, \epsilon_{\rm max} \lesssim 100$, as considered in this paper (with $\epsilon_{\rm max} = 10^{-5}$). The time scale $t_{\rm rad}(\gamma)$, together with the turbulent re-acceleration time scale and the escape time scale, is shown in Figue 4 as a function of the electron Lorentz factor $\gamma$ for the range of $B_{\rm gal} = 3-10$ $\mu$G (note that $t_{\rm iso}(\gamma)$, not shown in this figure, is much shorter than any other of the time scales plotted). As can be seen, $t_{\rm esc} \gg \max(t_{\rm acc}, \, t_{\rm rad})$ for all $\gamma \leq \gamma_{\rm br}$, and $t_{\rm acc} \gg t_{\rm rad}$ for $\gamma > \gamma_0$. All the electrons with Lorentz factors $\gamma < \gamma_{\rm KN} \equiv 1 / 4 \, \epsilon_{\rm max} \sim 10^{4.5}$ cool mainly via the inverse-Compton losses in the Thomson regime. Meanwhile, the electrons with $\gamma_{\rm KN} < \gamma < \gamma_{\rm cr}$, where $q \, F_{\rm KN}(\gamma_{\rm cr}) \equiv 1$, lose their energy mainly via the inverse-Compton emission in the Klein-Nishina regime. In the case of $B_{\rm gal} = 3$ $\mu$G, one has $\gamma_{\rm cr} \sim \gamma_{\rm br}$, while for $B_{\rm gal} = 10$ $\mu$G the critical electron Lorentz factor $\gamma_{\rm cr}$ is a factor of a few lower than that. Finally, the main process responsible for the cooling of the electrons with $\gamma > \gamma_{\rm cr}$ is synchrotron radiation.

\begin{figure}
\includegraphics[scale=1.50]{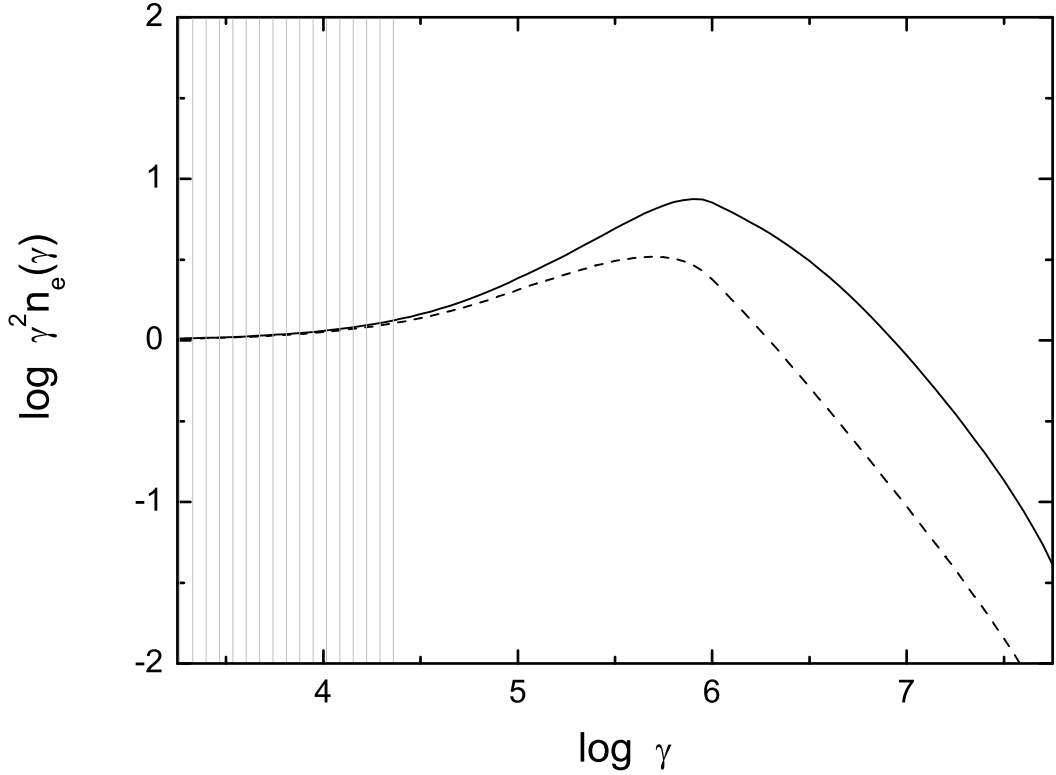}
\caption{Energy distribution of the electrons injected to the interstellar medium of the elliptical host, for the galactic magnetic field $3$ $\mu$G (solid line) and $10$ $\mu$G (dashed line), in normalized units. Shaded region indicate the electrons' energy range in which neglected re-acceleration effects are expected to become important.}
\end{figure}

The resulting electron energy distribution, $n_{\rm e}(\gamma)$, ignoring re-acceleration and escape effects, can be found from the continuity equation
\begin{equation}
{\partial n_{\rm e}(\gamma) \over \partial t} = {\partial \over \partial \gamma} \left\{ |\dot{\gamma}|_{\rm cool} \, n_{\rm e}(\gamma)\right\} + Q(\gamma) \quad ,
\end{equation}
\noindent
where $Q(\gamma)$ denotes injection of high-energy electrons through photon-photon annihilation, and $|\dot{\gamma}|_{\rm cool} = |\dot{\gamma}|_{\rm syn} + |\dot{\gamma}|_{\rm ic}$ is the total rate of the radiative cooling. The standard formulae give 
\begin{equation}
|\dot{\gamma}|_{\rm syn} = {4 \, c \sigma_{\rm T} \over 3 \, m_{\rm e} c^2} \, U_{\rm B} \, \gamma^2 \quad , \quad {\rm and} \quad |\dot{\gamma}|_{\rm ic} = {4 \, c \sigma_{\rm T} \over 3 \, m_{\rm e} c^2} \, U_{\rm B} \, \gamma^2 \, q \, F_{\rm KN}
\end{equation}
\noindent
\citep{mod05}. Note, that for the parameters considered here, and electron Lorentz factor $\gamma = 10^6$, the relative importance of the inverse-Compton and synchrotron energy losses, $|\dot{\gamma}|_{\rm ic} / |\dot{\gamma}|_{\rm syn} = q \, F_{\rm KN}$, is roughly $\sim 3$ for $B_{\rm gal} = 3$ $\mu$G, and $\sim 0.3$ for $10$ $\mu$G. Thus, the energy injected into the created pairs is re-radiated via their synchrotron and inverse-Compton (in the Klein-Nishina regime) processes in roughly comparable amounts. As for the function $Q(\gamma)$, we use approximation for the electrons freshly injected to the galactic volume $\cal{V}_{\rm gal}$ introduced by equation 22, to obtain the steady-state solution
\begin{equation}
n_{\rm e}(\gamma) = {3 \, m_{\rm e} c \over 4 \, \sigma_{\rm T}} \, {\int_{\gamma} d\gamma' \, Q(\gamma')\over \gamma^2 \, U_{\rm B} \, (1 + q \, F_{\rm KN})}  = {3 \, \tau \, L_{\rm inj} \over 4 \, c \, \sigma_{\rm T} \, \gamma_{\rm br}^2 \, {\cal{V}}_{\rm gal}} \, {\int_{\gamma} d\gamma' \, {\cal{I}}(\gamma') \over \gamma^2 \, U_{\rm B} \, (1 + q \, F_{\rm KN})} \quad .
\end{equation}
\noindent
This spectrum, in normalized units, is shown in Figure 5 for $B_{\rm gal} = 3$ $\mu$G (solid line) and $10$ $\mu$G (dashed line). The shaded region indicates the energy range for which the turbulent re-acceleration effects (neglected here) are expected to become important. As can be noted, at low electron energies $\gamma < \gamma_{\rm KN}$, for which the inverse-Compton cooling in the Thomson regime dominates, the energy spectrum has a standard form $n_{\rm e}(\gamma) \propto \gamma^{-p}$ with $p=2$, as expected in the case of the continuous injection of flat-spectrum ($Q(\gamma < \gamma_{\rm br}) \propto const$) particles, followed by the $|\dot{\gamma}|_{\rm ic/T} \propto \gamma^2$ energy losses. However, for $\gamma > \gamma_{\rm KN}$ the electron spectrum flattens, as a result of the dominant inverse-Compton/Klein-Nishina cooling \citep[see the most recent discussion in][and references therein]{mod05}, although in the case of  $B_{\rm gal} = 10$ $\mu$G this effect is relatively weak (the effective spectral index $p \sim 1.6$, to be compared with $p \sim 1.4$ for $B_{\rm gal} = 3$ $\mu$G). Meanwhile, at $\gamma > \gamma_{\rm cr}$ the electron sepectral index in both cases increases to $p = 3.5$, as expected for the dominant synchrotron cooling $|\dot{\gamma}|_{\rm syn} \propto \gamma^2$ of the continuously injected steep-spectrum ($Q(\gamma > \gamma_{\rm br}) \propto \gamma^{-2.5}$) electrons. 

\begin{figure}
\includegraphics[scale=1.50]{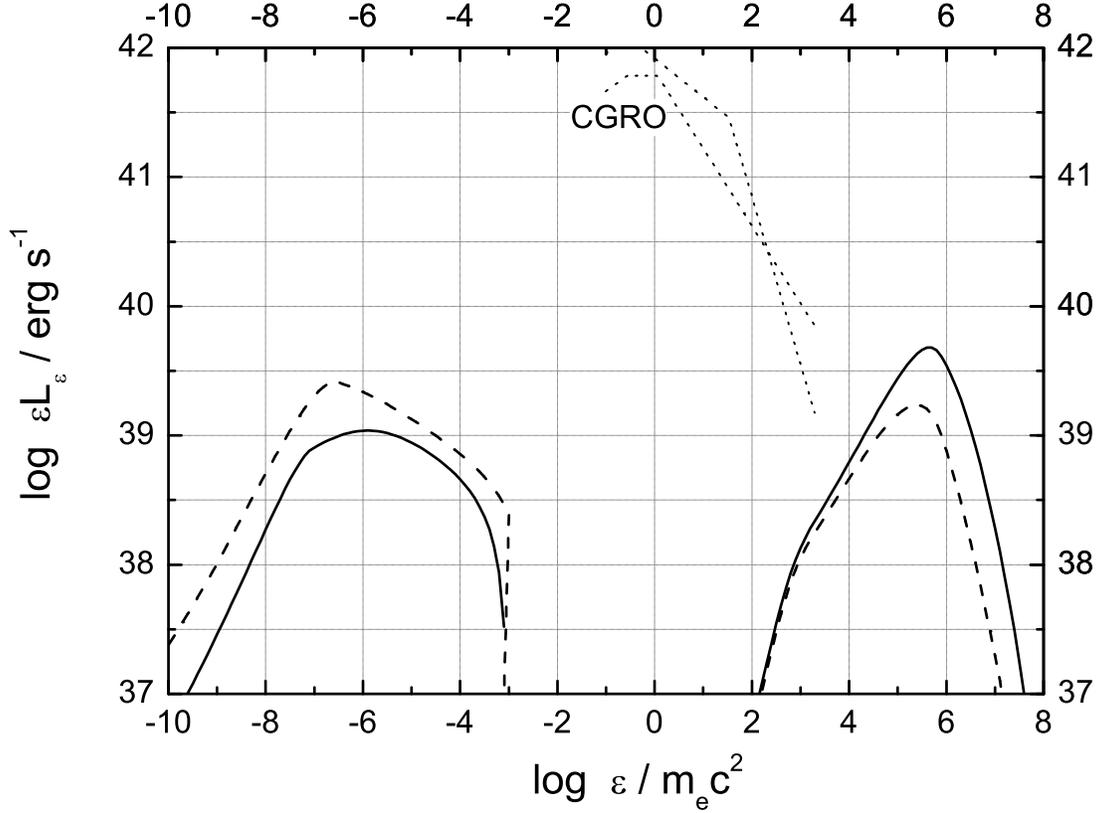}
\caption{Synchrotron and inverse-Compton spectra of the electrons injected into the interstellar medium of the elliptical host, for the galactic magnetic field $3$ $\mu$G (solid lines) and $10$ $\mu$G (dashed lines), and the injection luminosity $L_{\rm inj} = 10^{42}$ erg s$^{-1}$. Dotted lines indicates the $0.05-1000$ MeV spectrum of the Cen~A source as constrained by different instruments on board CGRO \citep[1991-1995; `low' and `intermediate' states;][]{ste98}.}
\end{figure}

With the evaluated electron energy distribution, one can find the energy spectrum of the resulting synchrotron and inverse-Compton emissions of the created pairs. In general, for the isotropic distribution of seed photons and particles, the respective luminosities can be simply written as $[\varepsilon L_{\varepsilon}]_{\rm syn/ic} = 4 \pi \, {\cal V}_{\rm gal} \, [\epsilon j_{\epsilon}]_{\rm syn/ic}$, where
\begin{equation}
[\epsilon j_{\epsilon}]_{\rm syn} = {1 \over 2} \, \left.{n_{\rm e}(\gamma) \, \gamma \over 4 \pi} \, |\dot{\gamma}|_{\rm syn} \, m_{\rm e} c^2\right|_{\gamma = \sqrt{(3/4) \, \epsilon_{\rm syn} \, (B_{\rm cr} / B)}} \quad ,
\end{equation}
\noindent
\citep[using $\delta$-approximation for the synchrotron emissivity; see][]{cru86}, $B_{\rm cr} \equiv 2 \pi \, m_{\rm e}^2 \, c^3 / h \, e = 4.4 \times 10^{13}$ G, and
\begin{equation}
[\epsilon j_{\epsilon}]_{\rm ic} = {3 \, \sigma_{\rm T} \, m_{\rm e} c^3 \over 16 \pi} \, \epsilon_{\rm ic}^2 \, \int d\gamma \, \int d\epsilon \, \, \, n_{\rm e}(\gamma) \, n_{\rm rad}(\epsilon) \, {{\cal F}_{\rm iso}(\gamma,\epsilon,\epsilon_{\rm ic}) \over \epsilon \, \gamma^{2}} \quad .
\end{equation}
\noindent
Here $\epsilon_{\rm syn/ic}$ denotes now the dimensionless energy of the synchrotron/inverse-Compton photons, electron energy distribution $n_{\rm e}(\gamma)$ is given by the equation 31, seed photons' spectrum
\begin{equation}
n_{\rm rad}(\epsilon) = {U_{\epsilon} \over \epsilon \, m_{\rm e} c^2} = {\langle U_{\rm rad}\rangle \over m_{\rm e} c^2 \, f_{\rm bol}} \, \epsilon^{-2} \, g(\epsilon)
\end{equation}
\noindent
is specified by the equation 5, and finally
\begin{equation}
{\cal F}_{\rm iso}(\gamma,\epsilon, \epsilon_{\rm ic}) = 2 \, {\cal P} \, \ln {\cal P} + {\cal P} + 1 - 2 {\cal P}^2 + {({\cal K} \, {\cal P})^2 \, (1 - {\cal P}) \over 2 \, (1 + {\cal K} \, {\cal P})}
\end{equation}
\noindent
with
\begin{equation}
{\cal K} \equiv 4 \, \epsilon \, \gamma \quad {\rm and} \quad {\cal P} \equiv {\epsilon_{\rm ic} \over 4 \, \epsilon \, \gamma \, ( \gamma - \epsilon_{\rm ic})} \quad ,
\end{equation}
\noindent
where $1/ 4 \gamma^2 \leq {\cal P} \leq 1$ \citep{blu70}. This leads to
\begin{equation}
[\epsilon L_{\epsilon}]_{\rm syn} = {\tau \, L_{\rm inj} \over 2 \, \gamma_{\rm br}^2} \, \left.{\gamma \, \int_{\gamma} d\gamma' \, {\cal{I}}(\gamma')  \over 1 + q \, F_{\rm KN}}\right|_{\gamma = \sqrt{(3/4) \, \epsilon_{\rm syn} \, (B_{\rm cr} / B)}} \quad ,
\end{equation}
\noindent
and
\begin{equation}
[\epsilon L_{\epsilon}]_{\rm ic} = {9 \, q \, \tau \, L_{\rm inj} \over 16 \, f_{\rm bol} \, \gamma_{\rm br}^2} \, \, \epsilon_{\rm ic}^2 \, \int d\gamma \, \int_{\max(\epsilon_{\rm min}, \epsilon_{\rm low})}^{\min(\epsilon_{\rm max},\epsilon_{\rm up})} d\epsilon \, \, \, {{\cal F}_{\rm iso}(\gamma,\epsilon,\epsilon_{\rm ic}) \, g(\epsilon) \over \epsilon^3 \, \gamma^4 \, (1 + q \, F_{\rm KN})} \, \int_{\gamma} d\gamma' \, {\cal{I}}(\gamma') \quad ,
\end{equation}
\noindent
where $\epsilon_{\rm low} \equiv \epsilon_{\rm ic}/4 \gamma \, (\gamma - \epsilon_{\rm ic})$, and $\epsilon_{\rm up} \equiv  \epsilon_{\rm ic} \gamma /(\gamma - \epsilon_{\rm ic})$. The evaluated luminosities are shown in Figure 6, for $L_{\rm inj} = 10^{42}$ erg s$^{-1}$ and the galactic magnetic fiel $B_{\rm gal} = 3$ $\mu$G (solid lines) and $10$ $\mu$G (dashed lines). In the case of the inverse-Compton emission, spectral index in the photon energy range $\epsilon_{\rm ic} = 10^{3} - 10^{6}$ (corresponding roughly to the electrons energies $\gamma_{\rm KN} < \gamma < \gamma_{\rm br}$), is $\alpha_{\rm ic/KN} \sim 0.36$ for $B_{\rm gal} = 3$ $\mu$G, and $\alpha_{\rm ic/KN} \sim 0.45$ for $10$ $\mu$G. These values are in agreement with the expected $\alpha_{\rm ic/KN} = p-1$ for the appropriate electron spectral index $p=1.4$ and $1.6$, respectively \citep[see a wide discussion in][]{mod05}. At $\epsilon_{\rm ic} > 10^6$, the inverse-Compton emission breaks to $\propto \epsilon_{\rm ic}^{-2.6}$, again as expected in the case of a steep power law electron continuum $\propto \gamma^{-3.5}$ at $\gamma > \gamma_{\rm br}$. In addition, Figure 6 presents the $0.05-1000$ MeV spectrum of the Cen~A source as constrained by different instruments on board CGRO \citep[1991-1995; `low' and `intermediate' states;][]{ste98}. As shown, the low-energy ($1-100$ GeV) flat-spectrum part of the predicted halo, if detected by the GLAST instrument in the future, could be spectraly distinguished from the (presumably nuclear) component observed by CGRO.

Figure 6 indicates that almost all the energy injected to the elliptical host via annihilation of the nuclear $\gamma$-ray emission on the starlight photon field is re-emitted via the synchrotron emission at $\sim 10^{13} - 10^{14}$ Hz frequencies, and via the inverse-Compton emission at $\gtrsim 0.1$ TeV photon energies. Unfortunatelly, the secondary synchrotron photons have almost the same energy as the target starlight photons, and at the same time much lower total luminosity, and therefore are not likely to be directly observed. However, the secondary $\gamma$-ray emission, although also relatively weak, is more promising for the detection. In particular, the analysis presented above indicates that at high photon energies ($> 0.1$ TeV) one should expect the photon flux from the galactic pair halo
\begin{equation}
F_{\rm iso} \sim {[\varepsilon_0 L_{\varepsilon_0}]_{\rm ic} \over 4 \pi \, d_{\rm L}^2 \, \varepsilon_0} \lesssim 10^{-11} \, \left({L_{\rm inj} \over 10^{42} \, {\rm erg/s}}\right) \quad {\rm ph \, cm^{-2} \, s^{-1}} \quad ,
\end{equation}
\noindent
with the Cen~A distance $d_{\rm L} = 3.4$ Mpc and $\varepsilon_0 = 0.2$ TeV. Note, that although the starlight radiation absorbs and reprocesses only a tiny ($\sim 1\%$) fraction of the nuclear $\gamma$-ray emisison, the small size of the resulting isotropic galactic pair halos ($\sim 100 \, r_{\rm b}$) make it quite easy to detect such structures. In fact, in the the case of Cen~A radio galaxy $100 \, r_{\rm b}$ corresponds to roughly $4$ kpc, or about $4$ arcmin, which is still within the active nucleus-centered flux integration region of the HESS telescope. Thus, the estimate given in equation 21 can be directly compared with the HESS upper limits given in \citet{aha05}. Such comparison implies that the time-averaged $\gamma$-ray output of the Cen~A nucleus is indeed $L_{\rm inj} < 10^{42}$ erg s$^{-1}$, which is already a meaningful result. This demonstrates an important aspect of the presented analysis: observations of nearby AGNs at TeV photon energies can provide important constraints on the time-averaged VHE $\gamma$-ray fluxes produced in their active centers, even if these sources are not belonging to the blazar sub-class (i.e., even if the nuclear portions of their jets are not inclined at small angles to the line of sight, and therefore their direct $\gamma$-ray emission is Doppler-hidden). Figure 7 shows in more details the expected $0.19-5$ TeV and $0.1-300$ GeV photon fluxes of the inverse-Compton emission for the discussed Cen~A galactic pair halo, together with the indicated HESS upper limit and the expected GLAST sensitivity \citep[$\sim 1.5 \times 10^{-9}$ ph cm$^{-2}$ s$^{-1}$ for one-year all-sky survey;][]{blo96}. As is shown, future observations by HESS and GLAST will be able to put independent constraints on both the low- and high-energy spectral parts of the predicted halo.

\begin{figure}
\includegraphics[scale=1.50]{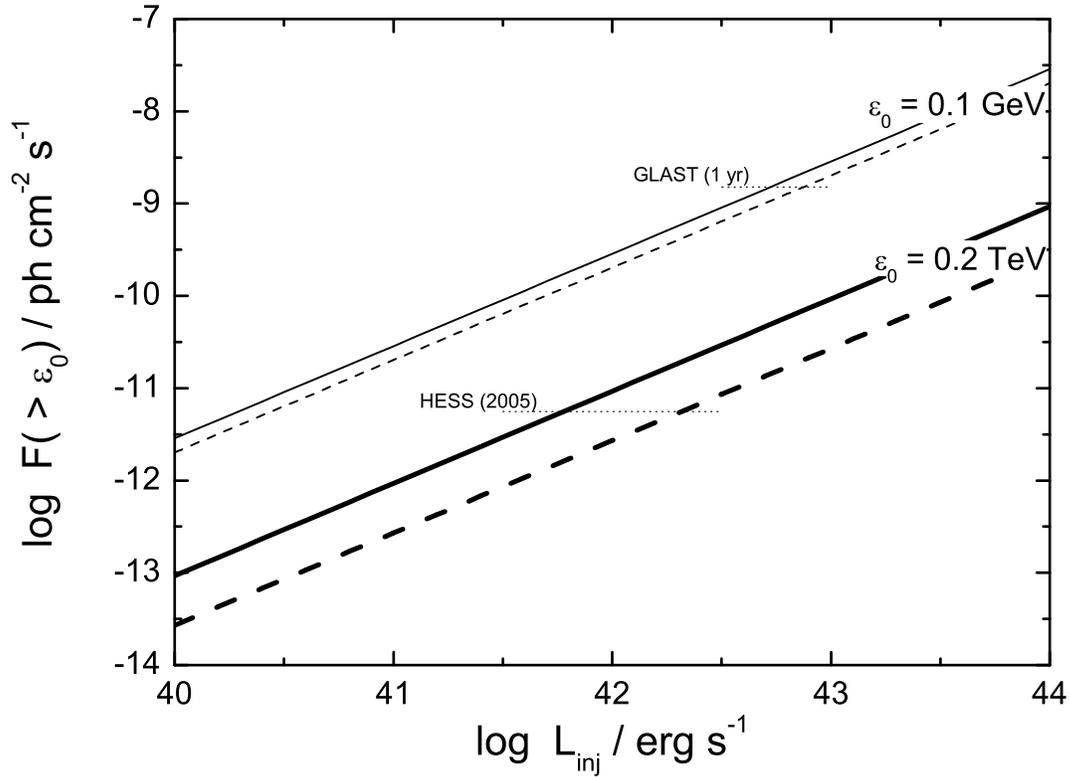}
\caption{The expected photon fluxes of the Cen~A halo for $\varepsilon_0 = 0.1$ GeV (thin lines) and $\varepsilon_0 = 0.2$ TeV (thick lines), corresponding to $B=3$ and $10$ $\mu$G (solid and dashed lines, respectively), as functions of the injection luminosity $L_{\rm inj}$. Horizontal dotted lines indicates HESS upper limit as given in \citet{aha05}, and the expected GLAST sensitivity \citep[$\sim 1.5 \times 10^{-9}$ ph cm$^{-2}$ s$^{-1}$ for one-year all-sky survey;][]{blo96}.}
\end{figure}

\section{Discussion and Conclusions}

Several BL Lac objects are confirmed sources of variable and strongly Doppler-boosted TeV emission produced in the nuclear portions of their relativistic jets. It is more than probable, that also many of the FR I radio galaxies, believed to be the parent population of BL Lacs, are TeV sources, for which strongly Doppler-hidden nuclear $\gamma$-ray radiation may be only too weak to be directly observed \citep[although see][for the case of nearby FR~I radio galaxy M~87 detected recently at TeV photon energies by HEGRA and HESS Cherenokov telescopes]{aha03,bei05}. Here we show, however, that about one percent of the total, time-averaged TeV radiation produced by the active nuclei of low-power FR~I radio sources is inevitably absorbed and re-processed by the photon-photon annihilation on the starlight photon field, and the following synchrotron/inverse-Compton emission of the created and quickly isotropized electron-positron pairs. Such a re-processed isotropic radiation could be detected in the cases of at least a few nearby FR~I radio galaxies, providing interesting constraints on the unknown parameters of the active nucleus and the elliptical host.

In the case of the Cen~A radio galaxy considered in this paper, we found that the discussed mechanism can give distinctive radiative features due to the isotropic $\gamma$-ray emission of the electron-positron pairs injected by the absorption process into the interstellar medium of the elliptical host (its inner parts in particular, roughly within the radius of $4$ kpc from the galactic center), and inverse-Compton upscattering thereby starlight radiation to the $\leq$ TeV photon energy range. The resulting $\gamma$-ray halo is expected to possess a spectral peak at $\sim 0.1$ TeV photon energies, preceded by a flat continuum due to the dominant Klein-Nishina cooling of the radiating electrons, and followed by a steep power-law $\propto \epsilon_{\rm ic}^{-(\Gamma_{\gamma}+0.5)}$, where $\Gamma_{\gamma}$ is the photon index of the primary (nuclear) $\gamma$-ray emission. Such a halo should be strong enough to be detected and mapped by stereoscopic systems of Cherenkov telescopes like HESS, and, at lower photon energies, by GLAST. 

All of the above findings should apply as well to the other nearby FR~I sources. We note in this context, that the kinetic power of the Cen~A jet, $L_{\rm j} \sim 10^{43}$ erg s$^{-1}$, is rather low when compared to other FR~I sources \citep[e.g., $L_{\rm j} \gtrsim 10^{44}$ erg s$^{-1}$ in the case of M~87 radio galaxy;][and references therein]{sta06b}. Thus, other (though more distant) objects of the FR~I type may posses more luminous isotropic halos than Cen~A analyzed here. Indeed, taking the time- and angle average nuclear $\gamma$-ray ($\sim$ TeV) luminosity $L_{\rm inj} \sim \eta_{\rm rad} \, L_{\rm j}$, where $\eta_{\rm rad}$ is the radiative efficiency, and the re-processed ($\gtrsim 0.1$ TeV) luminosity $L_{\rm iso} \sim \tau \, L_{\rm inj}$ with $\tau \sim 0.01$ as estimated in this paper, one can find that modern Cherenkov telescopes with the available sensitivity limit $10^{-13}$ erg cm$^{-2}$ s$^{-1}$ will be able to detect the discussed halos from the objects located within the radius rougly $d_{\rm L} \leq 100 \, \sqrt{(\eta_{\rm rad} / 0.1) \, (L_{\rm j} / 10^{44} \, {\rm erg \, s^{-1}})}$ Mpc. Is it therefore possible to attribute the $\gamma$-ray flux detected recently from M~87 system \citep{aha04} to the isotropic halo of its host galaxy? The answer is negative, since variability of this emission established on the time-scale of months and years \citep{bei05} excludes any extended $\gamma$-ray emission sites.

\appendix

\section[]{Opacity due to the `Dust Lane'}

\begin{figure}
\includegraphics[scale=1.50]{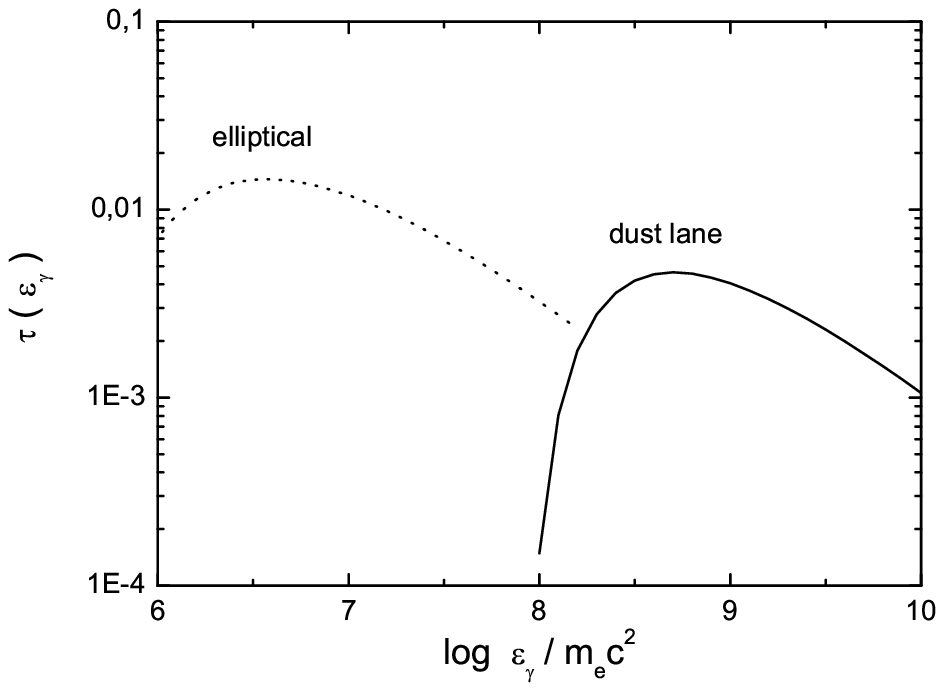}
\caption{Optical depth for photon-photo annihilation due to the dust lane (solid line) and the elliptical host (dotted line).}
\end{figure}

The continuum far-infrared emission of NGC~5128, produced most likely by massive young stars and diffuse cirrus clouds \citep{joy88,eck90}, is concentrated within the `dust lane'. We model this feature as a thin disc, perpendicular to the jet axis, centered on the active nucleus, and extending up to radii $R_{\rm fir} \sim 4$ kpc \citep[see][]{isr98}. We also restrict the analysis to $\lambda_{\rm fir} = 100$ $\mu$m radiation, for which we take the total observed (IRAS) flux $\sim 400$ Jy \citep{gol88}, corresponding to the luminosity $L_{\rm fir} \sim 1.6 \times 10^{43}$ erg s$^{-1}$. The number density of the far-infrared photons, assumed to be uniformly distributed within the `dust-lane', is then
\begin{equation}
n_{\rm fir, \epsilon}(\xi, \Omega) = {L_{\rm fir} \over 8 \, \pi^2 \, R_{\rm fir}^2 \, \epsilon_{\rm fir} \, m_{\rm e} c^3} \, \delta(\epsilon - \epsilon_{\rm fir}) \quad .
\end{equation}
\noindent
The appropriate optical depth for the photon-photon annihilation, analogous to the one given by equation 13, can be then evaluated as
\begin{equation}
\tau(\epsilon_{\gamma}) = {L_{\rm fir} \, r_{\rm b} \over 8 \, \pi^2 \, R_{\rm fir}^2 \, \epsilon_{\rm fir} \, m_{\rm e} c^3} \, \int_0^{\xi_{\rm t}} d \xi \, \int_{\cos[{\rm arccot}(\xi/100)]}^{+1} d\varpi \, \, \, (1 - \varpi) \, \, \, \sigma_{\gamma \gamma}(\epsilon_{\gamma}, \epsilon_{\rm fir}, \varpi) \quad .
\end{equation}
\noindent
This optical depth is shown on Figure 8 (solid line), together with the optical depth due to the starlight emission of the elliptical host evaluated peviously (dotted line). As shown, opacity due to the dust lane is much smaller than the opacity due to the elliptical host (because of the different spatial distribution of the target photons), and in addition regards only very high energies ($>100$ TeV) of the primary $\gamma$-rays.

\section*{Acknowledgments}
\L .S. and M.O. were supported by MEiN through the research project 1-P03D-003-29 in years 2005-2008. \L .S.\ acknowledges also the ENIGMA Network through the grant HPRN-CT-2002-00321, and thanks Wystan Benbow as well as the referee, Marcus B\"ottcher, for helpful comments.

\label{lastpage}


\begin{thebibliography}{}

\bibitem[Aharonian et al.(1983)]{aha83} Aharonian, F.A., Atoyan, A.M., \& Nagapetyan, A.M. 1983, Astrophysics 19, 187
\bibitem[Aharonian et al.(1994)]{aha94} Aharonian, F.A., Coppi, P.S., \& V\"olk, H.J. 1994, ApJ, 423, L5
\bibitem[Aharonian et al.(2003)]{aha03} Aharonian, F.A., et al. 2003, A\&A, 403, L1
\bibitem[Aharonian et al.(2004)]{aha04} Aharonian, F.A., et al. 2004, A\&A, 421, 529
\bibitem[Aharonian et al.(2005)]{aha05} Aharonian, F.A., et al. 2005, A\&A, 441, 465
\bibitem[Bailey et al.(1986)]{bai86} Bailey, J., Sparks, W.B., Hough, J.H., \& Axon, D.J. 1986, Nature, 322, 150
\bibitem[Beilicke et al.(2005)]{bei05} Beilicke, M., et al. 2005, In Proc. {\it `22nd Texas Symposium on Relativistic Astrophysics'}, 13-17 Dec. 2004, Palo Alto (USA)
\bibitem[Blandford \& Levinson(1995)]{bla95} Blandford, R.D., \& Levinson, A. 1995, ApJ, 441, 79
\bibitem[Bloom(1996)]{blo96} Bloom, E.D. 1996, Space Sci. Rev., 75, 109
\bibitem[Blumenthal \& Gould(1970)]{blu70} Blumenthal, G.R., \& Gould, R.J. 1970, Rev. Mod. Phys., 42, 237
\bibitem[Botti \& Abraham(1993)]{bot93} Botti, L.C.L., \& Abraham, Z. 1993, MNRAS, 264, 807
\bibitem[B\"ottcher \& Schlickeiser(1997)]{bot97} B\"ottcher, M., \& Schlickeiser, R. 1997, A\&A, 325, 866
\bibitem[B\"ottcher et al.(1997)]{boe97} B\"ottcher, M., Mause, H., \& Schlickeiser, R. 1997, A\&A, 324, 395
\bibitem[Burns et al.(1983)]{bur83} Burns, J.O., Feigelson, E.D., \& Schreier, E.J. 1983, ApJ, 273, 128
\bibitem[Capetti \& Balmaverde(2005)]{cap05} Capetti, A., \& Balmaverde, B. 2005, A\&A, 440, 73
\bibitem[Chiaberge et al.(2001)]{chi01} Chiaberge, M., Capetti, A., \& Celotti, A. 2001, MNRAS, 324, 33
\bibitem[Clarke et al.(1992)]{cla92} Clarke, D.A., Burns, J.O., \& Norman, M.L. 1992, ApJ, 395, 444
\bibitem[Colina \& de Juan(1995)]{col95} Colina, L., \& de Juan, L. 1995, ApJ, 448, 548
\bibitem[Coppi \& Blandford(1990)]{cop90} Coppi, P.S., \& Blandford, R.D. 1990, MNRAS 245, 453
\bibitem[Crusius \& Schlickeiser(1986)]{cru86} Crusius, A., \& Schlickeiser, R. 1986, A\&A, 164, 16
\bibitem[de Ruiter et al.(2005)]{rui05} de Ruiter, H.R., Parma, P., Capetti, A., Fanti, R., Morganti, R., \& Santantonio, L. 2005, A\&A, 439, 487
\bibitem[Dermer \& Schlickeiser(1994)]{der94} Dermer, C.D., \& Schlickeiser, R. 1994, ApJS, 90, 945
\bibitem[Eckart et al.(1990)]{eck90} Eckart, A., Cameron, M., Rothermel, H., Wild, W., Zinnecker, H., Rydbeck, G., Olberg, M., \& Wiklind, T. 1990, ApJ, 363, 451
\bibitem[Golombek et al.(1988)]{gol88} Golombek, D., Miley, G.K., \& Neugebauer, G. 1988, AJ, 95, 26
\bibitem[Gould \& Schr\'{e}der(1967)]{gou67} Gould, R.J., \& Schr\'{e}der, G.P. 1967, Phys. Rev. 155, 1404
\bibitem[Grindlay et al.(1975)]{gri75} Grindlay, J.E., Helmken, H.F., Brown, R.H., Davis, J., \& Allen, L.R. 1975, ApJ, 197, 9
\bibitem[Hardcastle et al.(2003)]{har03} Hardcastle, M.J., Worrall, D.M., Kraft, R.P., Forman, W.R., Jones, C., \& Murray, S.S., 2003, ApJ, 593, 169
\bibitem[Heidt et al.(2004)]{hei04} Heidt, J., Tr\"oller, M., Nilsson, K., J\"ager, K., Takalo, L., Rekola, R., \& Sillanp\"a\"a, A. 2004, A\&A, 418, 813
\bibitem[Israel(1998)]{isr98} Israel, F.P., 1998, A\&ARv, 8, 237
\bibitem[Jones et al.(1996)]{jon96} Jones, D.L., et al. 1996, ApJ, 466, 63
\bibitem[Joy et al.(1988)]{joy88} Joy, M., Lester, D.F., Harvey, P.M., \& Ellis, H.B. 1988, ApJ, 326, 662
\bibitem[Kataoka et al.(2006)]{kat06} Kataoka, J., Stawarz, \L ., Aharonian, F., Takahara, F., Ostrowski, M., \& Edwards, P.G. 2006, ApJ, 641, 158
\bibitem[Katarzy\'{n}ski et al.(2006)]{katar06} Katarzy\'nski, K., Ghisellini, G., Tavecchio, F., Gracia, J., Maraschi, L. 2006, MNRAS, 368, 52
\bibitem[Knapp et al.(1990)]{kna90} Knapp, G.R., Bies, W.E., \& van Gorkom, J.H. 1990, AJ, 99, 476
\bibitem[Kraft et al.(2003)]{kra03} Kraft, R.P., V\'azquez, S.E., Forman, W.R., Jones, C., Murray, S.S., Hardcastle, M.J., Worrall, D.M., \& Churazov, E., 2003, ApJ, 592, 129
\bibitem[Lauer et al.(1995)]{lau95} Lauer, T.R., Ajhar, E.A., Byun, Y.-I., Dressler, A., Faber, S.M., Grillmair, C., Kormendy, J., Richstone, D., \& Tremaine, S. 1995, AJ, 110, 2622
\bibitem[Marconi et al.(2006)]{mar06} Marconi, A., Pastorini, G., Pacini, F., Axon, D.J., Capetti, A., Macchetto, D., Koekemoer, A.M., \& Schreier, E.J. 2006, A\&A, 448, 921
\bibitem[Mathews \& Brighent(2003)]{mat03} Mathews, W.G., \& Brighent, F. 2003, ARA\&A, 41, 191
\bibitem[Moderski et al.(2005)]{mod05} Moderski, R., Sikora, M., Coppi, P.S., \& Aharonian, F. 2005, MNRAS, 363, 954
\bibitem[Morganti et al.(1991)]{mor91} Morganti, R., Robinson, A., Fosbury, R.A.E., di Serego Alighieri, S., Tadhunter, C.N., \& Malin, D.F. 1991, MNRAS, 249, 91
\bibitem[Moss \& Shukurov(1996)]{mos96} Moss, D., \& Shukurov, A., 1996, MNRAS, 279, 229
\bibitem[Rowell et al.(1999)]{row99} Rowell, G.P., et al. 1999, APh, 11, 217
\bibitem[Schlickeiser(1989)]{sch89} Schlickeiser, R. 1989, ApJ, 336, 243
\bibitem[Schlickeiser(2002)]{sch02} Schlickeiser, R. 2002, {\it `Cosmic Ray Astrophysics'}, Berlin: Springer
\bibitem[Sikora et al.(1994)]{sik94} Sikora, M., Begelman, M.C., \& Rees, M.J. 1994, ApJ, 421, 153
\bibitem[Sikora et al.(1997)]{sik97} Sikora, M., Madejski, G., Moderski, R., \& Poutanen, J. 1997, ApJ, 484, 108
\bibitem[Silva et al.(1998)]{sil98} Silva, L., Granato, G.L., Bressan, A., \& Danese, L. 1998, ApJ, 509, 103
\bibitem[Stawarz et al.(2003)]{sta03} Stawarz, \L ., Sikora, M., \& Ostrowski, M. 2003, ApJ, 597, 186
\bibitem[Stawarz et al.(2005)]{sta05} Stawarz, \L ., Siemiginowska, A., Ostrowski, M. \& Sikora, M. 2005, ApJ, 626, 120
\bibitem[Stawarz et al.(2006a)]{sta06a} Stawarz, \L ., Kneiske, T.M., \& Kataoka, J. 2006a, ApJ, 637, 693
\bibitem[Stawarz et al.(2006b)]{sta06b} Stawarz, \L ., Aharonian, F., Kataoka, J., Ostrowski, M., Siemiginowska, A., \& Sikora, M., 2006b, MNRAS, in press (astro-ph/0602220)
\bibitem[Steinle et al.(1998)]{ste98} Steinle, H., Bennett, K., Bloemen, H., Collmar, W., Diehl, R., Hermsen, W., Lichti, G.G., Morris, D., Schonfelder, V., Strong, A.W., \& Williams, O.R. 1998, A\&A, 330, 97
\bibitem[Tingay et al.(1998)]{tin98} Tingay, S.J., et al. 1998, AJ, 115, 960
\bibitem[Urry \& Padovani(1995)]{urr95} Urry, C.M., \& Padovani, P. 1995, PASP, 107, 803
\bibitem[Urry et al.(2000)]{urr00} Urry, C.M., Scarpa, R., O'Dowd, M., Falomo, R., Pesce, J.E., \& Treves, A. 2000, ApJ, 532, 816 
\bibitem[Wagner et al.(1995)]{wag95} Wagner, S.J., Camenzind, M., Dreissigacker, O., Borgeest, U., Britzen, S., Brinkmann, W., Hopp, U., Schramm, K.-J., \& von Linde, J. 1995, A\&A, 298, 688
\bibitem[Wang(2000)]{wan00} Wang, J.-M., 2000, ApJ, 538, 181
\end{thebibliography}
\end{document}